\documentclass[12pt]{article}
\usepackage{fullpage}           %Give a standard 1 inch margins on all sides    
\usepackage{amsmath,amssymb,float}
\usepackage{graphics}
\usepackage{subfigure}
\usepackage{epsfig}
\usepackage{lineno}
\begin{document}                                                                              
\title{Underlying SU(3) Symmetry of the Standard Model of Electroweak Interactions}                                                                              
\author{Martin A. Faessler \footnote{Supported by the DFG Cluster of Excellence
'Origin and Structure of the Universe'}\\
 Ludwig-Maximilians-University, Physics Department, 80799 Munich, Germany
 }
\date{\today}
\maketitle

%\newcommand{\leftexp}[2]{{\vphantom{#2}}^{#1}{#2}}
%Dirac Symbols                                                                  
\newcommand{\gz}{\gamma^0}
\newcommand{\gm}{\gamma^{\mu}}
\newcommand{\gf}{\gamma^5}                                               
\newcommand{\hp}{\tfrac{1+ \gamma^5}{2} }       
\newcommand{\hm}{\tfrac{1- \gamma^5}{2} }
\newcommand{\Lag}{\mathcal{L}}

\abstract{\noindent{
The Standard Model for electroweak interactions derives 
four vector gauge boson from an  SU(2)xU(1) symmetry. 
A doublet of  complex scalar (Higgs) bosons is added to generate
masses by spontaneous symmetry breaking. 
Both, the four vector  bosons and four scalar bosons 
corresponding to the Higgs  bosons of the Standard Model
are shown to emerge as gauge bosons 
from an underlying SU(3) symmetry of electroweak lepton interactions.
For the known leptons, the latter symmetry implies  
purely axial coupling of the Z boson to charged leptons,
if the photon is required to have  purely vector coupling. 
The corresponding Weinberg  angle is determined by sin$^2\theta = 1/4 $.
The same SU(3) symmetry  holds  for  quarks as well, if the observed   
fractional electric quark charges are averages of integer charges, 
according to the Han-Nambu scheme.         
}}
\\
\section{Introduction}
The Standard Model (SM) \cite{GSW,Textbooks} of electroweak interactions 
based on  SU(2) x U(1) symmetry 
describes the electroweak interactions of all observed 
elementary fermions, leptons and quarks, 
with all observed elementary bosons, vector and scalar fields.
There are  three (presently known)  generations  of fermions.  
They are distinct by flavor quantum numbers. 
The mixing of generations is an important issue of present particle physics.
However, for the following considerations it suffices to consider 
one generation of  leptons and quarks.
The three lepton states are, for the first generation:
A left handed neutrino, a left handed electron and a right handed electron.
For each of these   the corresponding antiparticle state exists:
Right handed antineutrino and positron  and  left handed antineutrino.
The short names to be used for the members of the triplets are:

leptons:   ($\nu_L, e_L,  e_R  $)  and antileptons:
($ \overline{\nu_L},  \overline{e_L},  \overline{\nu_R}  $) \\ 
The right-handed antineutrino  $\nu_R $ is not included in  the SM.
It is 'sterile' with respect to electroweak interactions. 
In the present work $\nu_R $ will be ignored, too.
Yet, recent  findings of neutrino oscillation experiments suggest  
it has to be included in a theory of all interactions of leptons.

The above triplets of  lepton and antilepton states  will be considered as 
fundamental representations of an electroweak  SU(3) symmetry.
In addition to the three leptons,  the four  quark states  of the first generation 
shall be   dealt with, two up (u) and two down (d) quarks and their antiparticles:

quarks:   ($ u_L, d_L,  u_R, d_R  $)  and antiquarks:
($ \overline{u_L},  \overline{d_L},  \overline{u_R}, \overline{d_R}  $) \\     
Before constructing  a Lagrange density which is  gauge  invariant    under  SU(3), 
a  heuristic  consideration is suggested. 
Assume a global SU(3) symmetry  
combining  the triplet of leptons   with that of  antileptons
to obtain a  regular octet and singlet representation. 
These combinations are considered to constitute 
(or to transmute into, or to be the decay products of) 
corresponding fields. 

In analogy to flavour-SU(3) \cite{Textbooks,Gellmann1964,Zweig1964,FritzschGellmann1971} 
and in concordance with the SM, 
the following quantum numbers are assigned to 
the three leptons (see Table \ref{tab2}): 
A weak isospin $I= 1/2 $ to the doublet of lefthanded leptons and 
no  weak isospin, $I= 0 $, and -if one wants- a  weak strangeness to the righthanded  electron.
Thus the triplet  ($\nu_L, e_L,  e_R  $)  corrresponds to the 
triplet  of up, down and strange quark,   ($ u, d, s  $),  
the basic representation of flavour SU(3) symmetry of light quarks.
A weak hypercharge 
$Y = 2(Q-I_3)$ can be defined based on the conserved charge $Q$ and  
weak isospin z- component  $I_3$. 
\begin{table}[htdp]
\centering
\begin{tabular}{   |p{2.2cm}|p{1.1cm}|p{1.3cm}|p{2.5cm}|p{1.5cm}    |p{2.5cm}| } \hline
  SM-fermion     &  $ Q$   &  $I_3$    &$Y=2(Q-I_3)$&$Q \cdot I_3$& $ \sum Q_iI^i_3/\sum Q^2_i$\\
\hline
    $ \nu _L$    & $\;\; 0$  &  $+1/2$   & $ -1$      & $\;\; 0$  &                  \\
    $ e_L $      & $  -1$    &  $-1/2$   & $ -1$      & $+1/2$    &                  \\
    $ e_R $      & $  -1$    & $\;\; 0$  & $ -2$      & $\;\; 0$  &  $ 1/4$       \\
    $ \nu _R$    & $\;\; 0$  & $\;\; 0$  &  $\;\; 0$  & $\;\; 0$  &                 \\
\hline
     $ u _L$     & $+ 2/3$  &  $ +1/2$  & $ +1/3 $    &  $+1/3$    &                 \\
     $ d_L $     & $ -1/3$  &  $ -1/2$  & $ +1/3 $    &  $+1/6$    &                \\
     $ u_R $     & $ +2/3 $ &  $\;\; 0$ & $ +4/3 $    &  $\;\; 0$   & $ 9/20$           \\
     $ d_R $     & $ -1/3$  &  $\;\; 0$ &   $ -2/3$   &  $\;\; 0$   &                 \\
\hline
\end{tabular}
\caption{SM electroweak quantum numbers of the first generation of leptons. 
Electric charge  $Q$  in units of $e$,
weak isospin z-component $I_3$  and weak hypercharge $Y= 2(Q-I_3) $.
The products $Q \cdot I_3 $  (one before last column) and the ratios
(last column) are relevant for the determination of   the Weinberg angle.
}
\label{tab2}
\end{table}

The peculiar feature of the basic triplet is that internal quantum numbers 
($ I_3, Y) $  are coupled to an external property, the chirality.
As a result, within  multiplets of higher dimensional representations there are 
particles with different spin.

\begin{figure}[t]
      \begin{center}
        \resizebox{0.5\columnwidth}{!}{\includegraphics{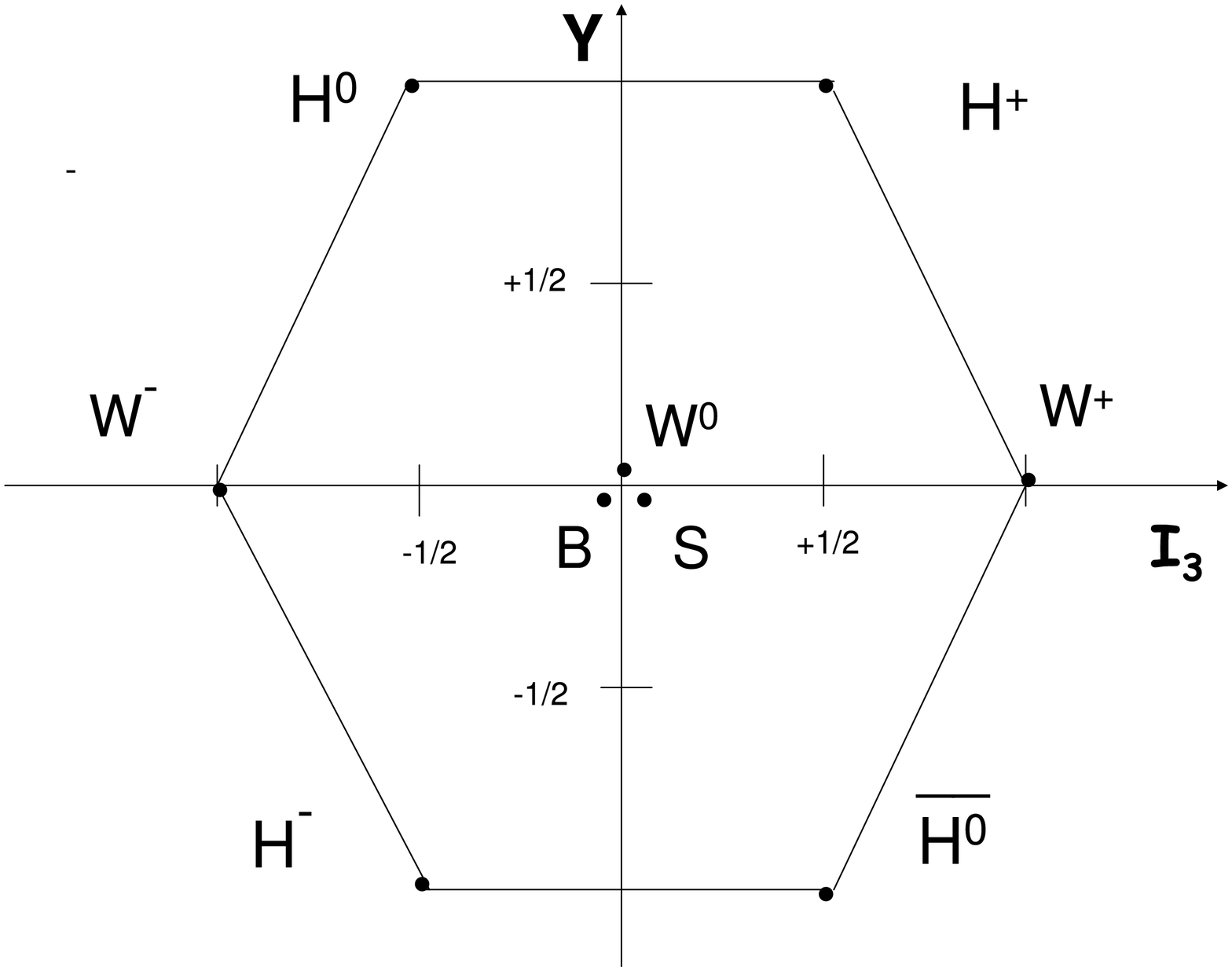} }  %0.5 instead of 0.\
%85
      \end{center}
\caption{Fields to be associated with
the regular octet and singlet representation of electroweak SU(3).
$Y$ and $I_3$ are the electroweak quantum numbers of the fields derived from those of their
'constituents', see Table \ref{tab2}. 
}
\label{fig1}
\end{figure}

Combining the three   leptons and their antileptons 
a nonet of bosons is obtained, see  Figure  \ref{fig1}.  
It contains a  weak isospin triplet of fields with $I_3 =-1,0,1 $. 
The spins of these fields are bound to be $S=1$ since 
the 'constituents' or decay products 
have opposite  chiralities.
(The same symbol has been used here for the antiparticles as will be used  for the adjoint 
bispinors lateron.  
Thus, if $\nu_L $  designates the annihilation of an (incoming) lefthanded  neutrino, 
$\overline{\nu_L}  $ designates the creation of an (outgoing) lefthanded neutrino, 
equivalent to an incoming righthanded antineutrino  $\overline{\nu}_R  $.)

These vector fields  are obvious candidates for  the triplet of weakly interacting vector bosons:\\

$ W^- = \overline{\nu_L} e_L, \;\;  \;\;
  W^0 = (\overline{\nu_L} \nu_L-\overline{e_L}e_L)/\sqrt{2}, \;\; \;\; 
  W^+ =   \overline{e_L}\nu_L $\\
\\
Two weak isospin  doublets with $I_3= -1/2, + 1/2 $  emerge from the combination
of leptons and antileptons with the same chirality.
Thus, the spins of the 'constituents'  are opposite  and  
their sum,  the spins of the  fields,  is $S=0$. 
They have  to  be  scalar or pseudoscalar  fields.\\

$H^-     =       \overline{\nu_L}e_R,   \;\;  \;\;
 H^+     =       \overline{e_R}\nu_L,   \;\;  \;\;
 H^0     =       \overline{e_R}e_L,     \;\;  \;\;
 \overline{H^0}=      \overline{e_L}e_R $\\
\\
It is suggestive to associate them with the weak isospin doublet 
of complex scalar Higgs fields assumed in the SM.

Finally, two more weak isoscalar $I = 0 $  vector fields are obtained. 
One of them is the isoscalar member of the octet, the other one the singlet representation.\\
 
$B_8 = (\overline{\nu_L} \nu_L+\overline{e_L}e_L  - 2 \overline{e_R}e_R )/\sqrt{6}  $

$S = ( \overline{\nu_L} \nu_L+\overline{e_L}e_L + \overline{e_R} e_R )/\sqrt{3}  $\\
\\
The $B_8 $  resembles the $B$ field associated with  U(1) of the SM 
which couples to the electroweak hypercharge.
However, the relative signs are different.
The different signs can be traced back to the
different signs of the strong hypercharges within the flavour nonet as compared to the weak 
hypercharges within the electroweak nonet.
As familiar from flavour-SU(3), the strange basic triplet member  has twice  the hypercharge 
with opposite sign as compared to the hypercharges of the  isospin doublet members up, down,
$(Y_d, Y_u, Y_s) = (1,1,-2)/3$.
However, the weak hypercharges of the leptons  all have the same sign, see Table \ref{tab2},
$(Y_{\nu_L}, Y_{e_L}, Y_{e_R}) = (-1,-1,-2)$.  

The same signs for the hypercharge currents interacting 
with  the $B$- field  shall be   needed  
in order to obtain the correct signs for the electric charge currents 
interacting with the $A$-field,
i.e. the photon, as will be shown below. 
This can be achieved  mixing (by field rotation)  the isoscalar octet member $ B_8 $ with the singlet $S$ as 
a first step. 
Thus, the singlet field and corresponding current  play a role in the present approach
whereas it is discarded in the  SM of electroweak interactions
as it is in the Standard Model  of  the strong interaction, QCD.
  
The next step after this rotation from $ B_8 $ and  $S$ to  $B$ and $S'$  
corresponds to the familiar standard mixing, 
i.e. rotation  by the Weinberg angle of the  resulting $B$ field   
and   of the $W^0$-field in order to obtain the  $A$- and  $Z$-boson.  
 
In the following, firstly  the  case  of leptons is considered. 
An attempt is made to derive both vector and scalar fields from a common local gauge symmetry.
This will require a somewhat different formulation of the Dirac equation.
The assumed SU(3) symmetry determines the Weinberg angle. 
It is shown,  that the same value of the Weinberg angle
emerges in the SM,  too,  if the world would consist only of the known 
leptons.
This is a consequence of   the choice of the hypercharge current 
attached  to  the  U(1)  symmetry  of the SM.

Subsequently, the SU(3) symmetry is extended to quarks.
It turns out that the Han-Nambu scheme \cite{HanNambu1965} explaining the fractional charges of quarks
as averages of three integer quark charges allows a straight-forward extension
of the SU(3) symmetry   to electroweak quark interactions.

%\newpage
\section{SU(3) Invariant  Lagrangian for Leptons}
The basic Dirac field operators  describing the leptons are 
4-component bispinors,  $\psi$ and expressed as column vector:
The adjoint bispinors are defined as $ \bar{\psi}= \psi ^{\dagger} \gz  $ and  row vectors.

The following covariant bilinear products, 
of the bispinors and their adjoints,
with  4x4 Dirac matrices, $\gamma^{\mu}$, where  $\mu=0,1,2,3 $
and their product $\gf= i \gamma^0 \gamma^1  \gamma^2  \gamma^3   $,  %.The products 
have specific behaviour (scalar, vector etc) under 
Lorentz transformations (boosts), rotations and space inversion:\\

$\bar{\psi} \cdot  \psi      \;\; \;\; \;\; $ scalar  
 
$\bar{\psi} \cdot \gf \cdot   \psi      \;\; \;\; \;\; $ pseudoscalar 

$\bar{\psi} \cdot \gm \cdot   \psi      \;\; \;\; \;\; $ vector 

$\bar{\psi} \cdot \gm \cdot \gf \cdot   \psi      \;\; \;\; \;\; $ axial vector \\ 
\\
The chirality projections, right (R)  and left (L)  handed  bispinors,  are obtained
by multiplication of the bispinor with the corresponding projection operators:\\

 $\psi_R = \tfrac{1}{2} (1+ \gf )\cdot \psi  \;\; \;\; \;\; $  and  
 $ \;\; \;\; \;\;\psi_L = \tfrac{1}{2} (1- \gf ) \cdot \psi$\\
\\
Some  properties of covariant bilinear products  of  chiral projections,
which are relevant for the present approach
are summarized here:\\

$ \overline{\psi} \cdot \psi_{R} = \overline{\psi_{L}} \cdot \psi_{R},   \;\;\;\;     $
$ \overline{\psi} \cdot \psi_{L} = \overline{\psi_{R}} \cdot \psi_{L},   \;\;\;\;    $
$ \overline{\psi} \cdot \gf \cdot \psi_{R} = \overline{\psi_{L}} \cdot  \gf \cdot \psi_{R} \;\;\;\;    $ 
$ \overline{\psi} \cdot \gf \cdot \psi_{L} = \overline{\psi_{R}} \cdot  \gf \cdot \psi_{L}   $ 

$ \overline{\psi} \cdot \gm  (\cdot \gf) \cdot \psi_{R} = \overline{\psi_{R}} \cdot  \gm  (\cdot \gf) \cdot \psi_{R} \;\;\;  $
and
$ \overline{\psi} \cdot \gm  (\cdot \gf) \cdot \psi_{L} = \overline{\psi_{L}} \cdot  \gm  (\cdot \gf)  \cdot \psi_{L}  $   \\
\\
In particular, the chiral projections  
turn out to be  orthogonal to their own adjoints  in the scalar and pseudoscalar product,
i.e. the scalar ($1$) and pseudoscalar ($\gf $) operators flip the chirality, 
whereas in case of the vector and axial vector currents the chirality is conserved 
and opposite chiralities are orthogonal:\\

$ \overline{\psi_R} \cdot{\psi_R}= \overline{\psi_L} \cdot \psi_L     
= \overline{\psi_R}  \cdot \gf  \cdot \psi_R= \overline{\psi_L} \cdot  \gf  \cdot \psi_L = 0 
\;\;\;\;\;\;\;\;\;\;  \;\;\;\;\;\;\;\;\;\;   \;\;\;\;\;\;\;\;\;\;  
\;\;\;\;\;\;\;\;\;\;  $ (1)\\  

$\overline{\psi_R}  \cdot \gm  \cdot \psi_L = \overline{\psi_L}  \cdot  \gm  \cdot \psi_R   
= \overline{\psi_R}  \cdot \gm  \cdot \gf  \cdot \psi_L = 
\overline{\psi_L}  \cdot \gm  \cdot \gf  \cdot \psi_R = 0
 \;\;\;\;\;\;\;\;\;\;  \;\;\;\;$ (2) \\
\\
%It has to be kept in mind  that chirality 
%is not necessarily conserved under Lorentz boosts except 
%in the limit of massless  fermions where chirality equals helicity.
%Helicity is a conserved quantum number.

The heuristic approach towards an SU(3) based electroweak interaction model 
for leptons outlined in the introduction
has shown that scalar fields and vector fields 
play a complementary role in the nine  fields 
coupling to the basic triplet of leptons.  
Thus a major challenge for the present approach  
has been  to justify the emergence of scalar  fields  in  addition to vector fields.
The solution is as follows  and first explained for the simpler case of U(1) and SU(2) symmetries 
if one considers a world of electrons only.
The  Lagrangian leading to the free Dirac equation via the Euler Lagrange equation is
\begin{eqnarray*}
  \Lag =  i  \bar{\psi} \partial_{\mu}  \gamma^{\mu} \psi  - \bar{\psi} m \psi  
\;\;\;\;\;\;\;\;\;\;  \;\;\;\;\;\;\;\;\;  \;\;\;\;\;\;\;\;\;\;   \;\;\;\;\;\;\;\;\; 
\;\;\;\;\;\;\;\;\;\;  \;\;\;\;\;\;\;\;\; \;\;\;\;\;\;\;\;\;\;  \;\;\;\;\;\;\;\;\;  
\label{freeDirace}   
\end{eqnarray*}
Gauge symmetry under the U(1)  transformations\\

$ U =   e^{ i e \alpha(x)}  \;\;\;\;\;\; $
so that 
$ \;\;\; \psi(x)  \rightarrow  e^{ i e \alpha(x)} \psi(x) \;\;\;\;\;\;  $  and 
$ \;\;\;  \overline{\psi(x)}  \rightarrow  e^{ -i e \alpha(x)}   \overline{\psi(x)} $\\
\\
requires the existence of  
a vector field  potential $A_{\mu} $ (naturally to be associated with  the photon).
This field must transform under  U(1) transformations as:\\

$ A_{\mu} \rightarrow A_{\mu} + \partial_{\mu} \alpha(x) $\\
\\ 
It is proposed to rewrite the free Dirac Lagrangian in the following way: 
\begin{eqnarray*}\label{freeDirac2} 
    \Lag =  i  \bar{\psi} \partial_{\mu}  \gamma^{\mu} \psi  -i \bar{\psi} d_{\tau}  \psi  
\;\;\;\;\;\;\;\;\;\;\; \;\;\;\; \;\;\;\;\;\;\;\;\;  \;\;\;\;\;\;\;\;\; 
\;\;\;\;\;\;\;\;\;\;  \;\;\;\;\;\;\;\;\; \;\;\;\;\;\;\;\;\;\;  
\;\;\;\;\;\;\;\;\;\;\;\;\;\;\;\;\;\; \;\;\;\;\;\;\;\;\;  (3)  
\end{eqnarray*}  
where the  derivative replacing the mass m in the standard Lagrangian, 
$i  d_{\tau} = i  d/d \tau =  i  u^{\mu}  \partial_{\mu}  $,  is taken 
with respect to the eigentime  $\tau$ of the lepton and  
$ u^{\mu} $ is the four-velocity of the lepton. 
It implements the correspondence between classic particle 4-momentum
and quantum operators $p_{\mu}   \rightarrow  i\partial_{\mu} $  by\\ 

$ m   \rightarrow  i d_{\tau}: = i  u^{\mu}  \partial_{\mu}   $\\ 
\\
With this modification, invariance under the above  unitary transformation U 
requires the addition of   the scalar product
of the vector field potential to the lepton 4-velocity  $A = u^{\mu} A_{\mu}$.
This may be interpreted  as an effectively scalar field. 
The  Lagrangian becomes, with  the standard definition and coupling of the 
photon field  $A_{\mu} $ 
and the standard replacement of the covariant derivative  $\partial_{\mu} $ by
$   D_{\mu} =  \partial_{\mu} -ie A_{\mu}  $:
\begin{eqnarray*}
    \Lag =  \bar{\psi} \gamma^{\mu} (i  \partial_{\mu} + e A_{\mu} ) \psi  -  \bar{\psi} u^{\mu}  (i \partial_{\mu} +e A_{\mu}  )  \psi  \;\;\;\;\;\;\;\;\;\; 
\;\;\;\;\;\;\;\;\;\;  \;\;\;\;\;\;\;\;\;\; \;\;\;\;  \;\;\;\;\;\;\;\;\;\;  \;\;\;\;\;\;\;\;\;\;\;\;\;\;\;\;\;\;  (4) 
\label{Dirac2} %\tag{5}
\end{eqnarray*}     
This Lagrangian is invariant  
under the U(1) transformation defined above.

An  instructive intermediate step is to 
extend the U(1) symmetry to SU(2), taking as a basis the  two chiral 
projections $ (e_R, e_L)$ of the electron defined above.
This is one of the subgroups  of the SU(3) considered in the present approach. 
Gauge symmetry with respect to the  unitary transformations\\
 
$U=  e^{  i e \alpha_0 (x) \cdot 1  +i e \overrightarrow{\alpha}(x) \cdot \vec{\tau} } $\\
\\
where $ \vec{\tau} $   contains the three Pauli matrices and the $1 $ behind $ \alpha_0 (x) $ 
is the 2x2  unity  matrix,   
leads to  two vector and two scalar fields and
the covariant derivative including minimal coupling is
$   D_{\mu} =  \partial_{\mu} -ie A^0_{\mu} -ie  \vec{\tau} \overrightarrow{A} $. 
Note that  here, for SU(2),
the singlet representation  is included by means of the first term, 
as it shall be done  further below for SU(3).  
The corresponding vector potential $ A^0_{\mu}  $ 
can be interpreted as being the photon, since it turns out to have  
a purely vector coupling to the electron, i.e.  couples to the vector current 
$  \sim (\overline{e_L} \gamma^{\mu} e_L   + \overline{e_R}    \gamma^{\mu} e_R )     $. 
An  other one associated  with the  phase transformations $ \alpha_3 (x) \cdot \tau_3    $  
couples to the electron like an axial vector field, 
i.e. to the axial current $ \sim (\overline{e_L} \gamma^{\mu} e_L   -\overline{e_R} \gamma^{\mu} e_R )  $. 
It is suggestive to associate it with the Z-boson.
The two scalar fields emerging from 
the two phase changes  $ \alpha_1 (x) \tau _1$ and   $ \alpha _2 (x) \tau _2$
connect the left handed electron with the adjoint of the right 
handed electron and vice versa.  
These two scalar fields  are the candidates to be   
responsible for the mass of the electron.
Define the operator and its adjoint for  the doublet of lepton fields by
\begin{eqnarray*}\label{doublet}
  \phi =
           \begin{pmatrix}
              e_L    \\
              e _R   
            \end{pmatrix} \;\; \;\; \;\; \;\; \;\; \;\;
\bar{\phi}  =  (  \overline{e_L},  \overline{e_R}) 
 \;\; \;\; \;\; \;\; \;\; \;\; \;\; \;\; \;\; \;\; \;\; \;\; \;\; \;\; \;\; \;\; \;\; \;\; 
\end{eqnarray*}
And note that the 'mass'  term of the free Lagrangian drops out: 
$ -i\bar{ \phi} d_{\tau}  \phi  =   -i \overline{ e_R} d_{\tau} e_R - i \overline{ e_L} d_{\tau} e_L  = 0$
owing to the orthogonality of 
the chirality eigenstates with their adjoints, see equation (1). 
But the mass term is replaced by the two interaction terms with the scalar fields: 
$- e\bar{\phi} u^{\mu}  (\tau _1   A_{\mu}^1+ \tau _2   A_{\mu}^2  )  \phi  $.
The effectively scalar fields may serve  for the mass generation by spontaneous symmetry breaking.
 
The SU(2) symmetry of the SM corresponds to another subgroup of   SU(3).  
The basic leptons are the two left handed particles, $\nu_L, e_L$.
Four  vector fields -and no scalar field  emerge 
if one includes phase transformation $ \alpha_0 (x) $ as for the SU(2) above.
The corresponding  current corresponding to the singlet representation  
is  
$  J^{\mu}_0 \; \sim \;    (\bar{\nu _L} \gamma^{\mu} \nu _L 
                         +  \bar{e _L} \gamma^{\mu} e _L ) $, 
which is  part of the current introduced through U(1) of the SM.  
In other words, the Standard Model does not directly  include this  term 
but the  separate U(1) symmetry with respect to the hypercharge current 
$  J^{\mu}_Y \sim  (J^{\mu}_0  + 2   \bar{e _R} \gamma^{\mu} e _R ) $ contains part of it.   
The three vector fields required by  SU(2) of the SM and SU(2) subgroup of the presented
SU(3) model are the $W$ bosons.
The vector field associated with the U(1) symmetry and the current 
$  J^{\mu}_Y  $ 
of the SM  is the $B$ boson.

In the present work  the basic  triplet  of SU(3) consists of  the  leptons.
They are contained in the column  three vector 
called  $ \phi $ and its adjoint:  
\begin{eqnarray*}\label{triplet}
  \phi =
           \begin{pmatrix}
              \nu_L   \\
              e_L    \\
              e _R   
            \end{pmatrix} \;\; \;\; \;\; \;\; \;\; \;\;
\bar{\phi}  =  ( \overline{\nu_L},  \overline{e_L},  \overline{e_R}) 
 \;\; \;\; \;\; \;\; \;\; \;\; \;\; \;\; \;\; \;\; \;\; \;\; \;\; \;\; \;\; \;\; \;\; \;\; 
\end{eqnarray*}
The free Lagrangian is obtained from the Lagrangian, see equation (3), 
replacing the single Dirac field operator $\psi $ by the 3- vector 
of Dirac field operators  $\phi $. 
SU(3) gauge symmetry is assumed, i.e. invariance of the Lagrangian with respect to 
the special unitary transformation\\

$\;\; \;\; \;\; \;\; U=  e^{ i g \overrightarrow{\alpha}(x) \cdot \vec{\lambda}}  
\;\; \;\; \;\; \;\; \;\;\; \;\; \;\; \;\; \;\; \;\; \;\; \;\; \;\; \;\; \;\; \;\; \;\; \;\; 
\;\; \;\; \;\; \;\; \; \;\; \;\; \;\; \;\; \;\;\;\; \;\; \;\; \;\; \;\; \;\; \;\; \;\; 
 \;\; \;\; \;\; \;\; \;\; \;\; \;\; \;\;  \;\; \;\; \;\; \;\; (5) $\\ 
\\
The 9-vector  $ \vec \lambda = (\lambda_0, \lambda_1,..., \lambda_8) $ 
contains the standard Gell-Mann matrices ($ \lambda_1,..., \lambda_8 $) %(Close, page30), 
implemented by the 
unity matrix $\lambda_0  = 1 \cdot \sqrt{2/3}$ needed for the SU(3) singlet term.
(The factor  $\sqrt{2/3} $ has to be applied %to the  unity matrix   %$1$ 
in order to have  the same normalisation 
as the other matrices.) 
Requiring gauge symmetry with respect to  U 
and minimal coupling will lead to interaction terms with vector and scalar 
fields, depending on the combination of left and right handed leptons as 
dictated by the $ \lambda$ matrices.
The  initial names to be used for  the nine   fields to be expected  are 
$ X_{\mu}^0,...,X_{\mu}^8 $. %  and    $  X^0,...,X^8  $.  
The Lagrangian implemented by the interaction term 
(but not yet including  the field energy density terms) is now:  
\begin{eqnarray*}% \label{Dirac2}
   \Lag =  \bar{\phi} \gamma^{\mu} (i \partial_{\mu} + g\vec{\lambda} \vec{X}_{\mu} ) \phi  
     -  \bar{\phi}      u^{\mu} (i \partial_{\mu} + g\vec{\lambda} \vec{X}_{\mu} ) \phi
     = \bar{\phi} \gamma^{\mu} i \partial_{\mu}  \phi 
               + g\bar{\phi} ( \gamma^{\mu}- u^{\mu} )\vec{\lambda}\vec{X}_{\mu} \phi  
\;\;\;\;\;\;\;\;\;\;  (6)
\label{Dirac2} %\tag{5}
\end{eqnarray*}
This Lagrangian is invariant  under the SU(3) transformations (5).
Due to the orthogonality properties, see equations  (2), of the chiral  projections, 
the  term  
$  -  \bar{\phi}      u^{\mu} (i \partial_{\mu}  ) \phi  $
of the free 
Lagrangian has dropped out for the present choice 
of the basic triplet- as it  already did for the extensions to SU(2).
Thus the masses have to arise from the interaction of the leptons 
with the effectively  scalar fields
$u^{\mu}   \vec{\lambda} \vec{X}_{\mu} $. 

The  interaction term written as an explicit sum  is \\ 

$  g  \bar{\phi} (\gamma^{\mu} - u^{\mu} ) \vec \lambda   \overrightarrow{X_{\mu} }  \phi 
  =  \sum _{i=0,.,9}  g  \bar{\phi} (\gamma^{\mu}  - u^{\mu})  \lambda_i  X_{\mu}^i \phi
  =: \sum _{i=0,.,9}  \Lambda _i 
\;\; \;\;  \;\; \;\; \;\; \;\; \; \;\; \;\; \;\; \;\; \;\; \;\; \;\; (7)$\\
\\
Considering the vector current interaction terms, 
$ \Lambda _i  =  
g \cdot \bar{\phi}\cdot \gamma^{\mu} \cdot  \lambda_i  \cdot X_{\mu}^i    \cdot \phi$,
it is obvious from the orthogonality relations (2) that only five of them survive,
namely those with chirality  conserving currents.
For the scalar interaction  terms, 
$ \Lambda _i  =  
g \cdot \bar{\phi} \cdot u^{\mu} \cdot  \lambda_i  \cdot X_{\mu}^i    \cdot \phi$,
the other four  with  chirality changing currents survive. 
The complete list is:\\   
\\
$ \Lambda _0   
  = g  \cdot (\overline{\nu_L}  \gamma^{\mu} \nu_L + \overline{e_L} \gamma^{\mu} e_L 
           + \overline{e_R}    \gamma^{\mu} e_R )         \cdot  X_{\mu}^0 \cdot \sqrt{2/3}$\\
$ \Lambda _1 
  = g  \cdot (\overline{\nu_L}   \gamma^{\mu} e_L   + \overline{e_L} \gamma^{\mu} \nu_L ) \cdot X_{\mu}^1  $\\ 
$ \Lambda _2      
  = g  \cdot (-i \overline{\nu_L} \gamma^{\mu} e_L  + i\overline{e_L} \gamma^{\mu} \nu_L) \cdot X_{\mu}^2  $\\
$ \Lambda _3    
  = g  \cdot (\overline{\nu_L}   \gamma^{\mu} \nu_L -  \overline{e_L} \gamma^{\mu} e_L)   \cdot  X_{\mu}^3 $\\
$ \Lambda _4    
  =  g \cdot (\overline{\nu_L}    u^{\mu}      e_R +  \overline{e_R}  u^{\mu} \nu_L)      \cdot  X_{\mu}^4 
\;\; \;\; \;\; \;\; \; \;\; \;\; \;\; \;\; \;\; \;\; \;\;\;\; 
\;\;\;\;\;\;\;\;\;\;  \;\;\; \;\;\;\;\;\;\;\;\;\;   \;\; \;\; \;\; \; \;\; \;\; \;\; \;\; \;\; \;\; \;\;  \;\;\; \;\;\;\;\;\;\;\; (8)$\\
$ \Lambda _5     
  = g  \cdot (-i \overline{\nu_L} u^{\mu}      e_R + i\overline{e_R}  u^{\mu} \nu_L)      \cdot  X_{\mu}^5 $ \\
$ \Lambda _6     
  = g  \cdot (\overline{e_L}      u^{\mu}      e_R +  \overline{e_R}  u^{\mu}  e_L)       \cdot  X_{\mu}^6 $ \\
$ \Lambda _7     
  = g  \cdot (-i \overline{e_L}   u^{\mu}      e_R +  i\overline{e_R} u^{\mu}  e_L)       \cdot  X_{\mu}^7 $ \\
$ \Lambda _8   
  = g  \cdot (\overline{\nu_L}  \gamma^{\mu} \nu_L +  \overline{e_L}  \gamma^{\mu} e_L -
                             2 \overline{e_R} \gamma^{\mu} e_R )                  \cdot  X_{\mu}^8 /\sqrt{3}$\\ 
\\
The interaction terms will be discussed in the following with a focus on the comparison with
the SM and on the determination of the Weinberg angle.

\section{Comparison of  Interaction Terms}
Recomposing the first two terms  by replacement of  $  \lambda_1,  \lambda_2$  
and  $ X_{\mu}^1, X_{\mu}^2 $ with
$\lambda_{1\pm 2}  = \lambda_1 \pm i \lambda_2 $  and
$    X_{\mu}^{1\pm 2} = (X_{\mu}^1 \pm i  X_{\mu}^2)/ \sqrt{2} $
the  sum  of the first two terms $\Lambda _1,  \Lambda_2 $
is transformed into a new sum
$ \Lambda _{1 + 2} + \Lambda _{1 - 2} = \Lambda _1 + \Lambda _2 $ with:\\
\\
$ \Lambda _{1+2}
= (g / \sqrt{2}) \cdot \bar{\phi} \cdot \lambda_{1+2}  \cdot \gamma^{\mu}  \cdot \phi \cdot  X_{\mu}^{1-2}
= (\sqrt{2}g)    \cdot  \overline{\nu_L}  \gamma^{\mu}  e_L \cdot  X_{\mu}^{1-2}$\\
$ \Lambda _{1-2}
= (g / \sqrt{2}) \cdot \bar{\phi} \cdot \lambda_{1-2}  \cdot \gamma^{\mu}  \cdot \phi \cdot  X_{\mu}^{1+2}
= (\sqrt{2}g)   \cdot   \overline{e_L}  \gamma^{\mu}  \nu_L \cdot  X_{\mu}^{1+2}$\\
\\
Next, it is  shown that  the  three terms 
$  \Lambda _{1+2}, \Lambda _{1-2}, \Lambda _3 $
are proportional to   the interaction terms
of the weak charged and neutral currents with the W- bosons 
of the SU(2) part of the Standard Model (SM).
This is  to be expected since  the matrices $\lambda_1, \lambda_2, \lambda_3 $
project the electroweak lefthanded lepton doublet
out of the present triplet and   their effective part is a copy 
of  the Pauli matrices  $\tau_1, \tau_2, \tau_3 $.

The interaction term of the SM Lagrangian  for the SU(2) symmetric part is, 
before the rotation by the Weinberg angle,\\
\\
$  \Lambda _{SU(2)}^{SM}  =
g_{SM} \cdot \overline{\chi_L}\cdot  \gamma^{\mu} \cdot  \vec T  \cdot  \overrightarrow{W_{\mu}} \cdot \chi_L$
$  =g_{SM}  \overrightarrow{J^{\mu}} \overrightarrow{W_{\mu}}$\\
\\
Here, $\chi_L$  is the column vector of the (weak isospin)  doublet
of the two left-handed bispinors associated with   $\nu_L$ and $e_L$  and
$ \overline{\chi_L} $ is the row vector  adjoint  to  $\chi_L$:
\begin{eqnarray*}\label{doublet}
  \chi_L =
           \begin{pmatrix}
              \nu_L   \\
               e_L
            \end{pmatrix}  \;\; \;\; \;\; \;\; \;\; \;\;
   \overline{\chi_L }  =  ( \overline{\nu_L},  \overline{e_L})
\end{eqnarray*}
The 3-component vector $ \vec T = \vec \tau /2 $  
contains the three generators of weak isospin.
The vector $  \overrightarrow{W_{\mu}} $ represents the three associated vector fields.
The three weak isospin currents $ \overrightarrow{J^{\mu}} $ contained in the
SU(2) interaction term above are:\\

$[ J^{\mu}_1, J^{\mu}_2,J^{\mu}_3]  = \tfrac{1}{2}[  (\overline{\nu_L}\gamma^{\mu} e_L + \overline{e_L}\gamma^{\mu} \nu_L),
                                                  (-i\overline{\nu_L}\gamma^{\mu} e_L + i\overline{e_L}\gamma^{\mu} \nu_L),
                                                 (\overline{\nu_L}\gamma^{\mu} \nu_L - \overline{e_L}\gamma^{\mu} e_L)]   $\\
\\
In the SM, the first two interaction terms are recomposed 
into two charged current interactions, 
as in the present approach, which was guided by the SM.
The two charged currents are:\\

$ J^{\mu}_+ := J^{\mu}_1 + i J^{\mu}_2 =   \overline{\nu_L} \gamma^{\mu} e_L  \;\;\;\;\;\;\;\;\;\;\;\;\;\;$
and
$ \;\;\;\;\;\;\;\;\;\;\;\;\;\;\;\;\;\;\;\;
J^{\mu}_- := J^{\mu}_1 - i J^{\mu}_2 =   \overline{e_L}\gamma^{\mu} \nu_L   $\\
\\
The two associated charged fields are defined  as were the fields
$X_{\mu}^{1-2}, X_{\mu}^{1+2}  $:\\

$W_{\mu}^+ = \tfrac{1}{\sqrt{2}} (W_{\mu}^1 -iW_{\mu}^2) \;\;\;\;\;\;\;\;\;\;\;\;\;\;\;\;\;\;\;\;$
and
$ \;\;\;\;\;\;\;\;\;\;\;\;\;\;\;\;\;\;\;\; W_{\mu}^- =   \tfrac{1}{\sqrt{2}} (W_{\mu}^1 +iW_{\mu}^2)$\\
\\
Now, it is obvious that the three interaction terms of SU(2) of the SM:\\
\\
$  [\Lambda _-^{SM},\; \Lambda _+^{SM},\;  \Lambda _3^{SM}] $
 $ =  g_{SM}  [\tfrac{1}{\sqrt{2}}  \overline{\nu_L}  \gamma^{\mu}  e_L    W_{\mu}^{+}, \;
             \tfrac{1}{\sqrt{2}}   \overline{e_L}   \gamma^{\mu}  \nu_L   W_{\mu}^{-} ,  \;
           \tfrac{1}{2}    (\overline{\nu_L} \gamma^{\mu} \nu_L - \overline{e_L} \gamma^{\mu} e_L)   W_{\mu}^0]$\\
\\
and those  of the present SU(3) approach,  $  \Lambda _{1+2}, \Lambda _{1-2}, \Lambda _3 $,
are consistent, apart from the fact that
the  coupling constants are  differently defined in the SM.
(There is a factor 2 between the two definitions.
This can be traced back to the different normalisations of the $T$ 
used in the SM  and of  the $\lambda$ matrices used here 
for  SU(3) transformations, equation (5).)

The  interaction term for the U(1) part of the SM describes the interaction of
a weak hypercharge current with a vector field called $B_{\mu}$.
The hypercharge current of the SM is  due to the weak hypercharge $Y= 2(Q-I_3)$
and is therefore defined as:\\
\\
$J^{\mu}_Y  = 2 J^{\mu}_{e.m.} - 2 J^{\mu}_3
 = (-2) (\overline{e_L} \gamma^{\mu} e_L + \overline{e_R} \gamma^{\mu} e_R) -
        2 \cdot \tfrac{1}{2} (\overline{\nu_L} \gamma^{\mu}  \nu_L - \overline{e_L}  \gamma^{\mu} e_L)$\\
\\
The corresponding  interaction term is:\\
\\
$ \Lambda _{U(1)}^{SM} = \tfrac {g'_{SM}}{ 2}  J_Y^{\mu} B_{\mu}
    = \tfrac {g'_{SM}}{ 2} ( -\overline{\nu_L} \gamma^{\mu}  \nu_L    -   \overline{e_L}  \gamma^{\mu} e_L
      - 2\overline{e_R} \gamma^{\mu} e_R )   B_{\mu}$\\
\\
The term $\Lambda _{U(1)}^{SM} $ resembles the term $\Lambda _8 $.
However, as already mentioned, the relative signs in   $   \Lambda_8  $
do not correspond to those in $\Lambda _{U(1)}^{SM} $.
An unavoidable  consequence is that after mixing of the
hypercharge candidate vector field  $  X_{\mu}^8 $ 
with  $ X_ {\mu}^3 = W_{\mu}^0$  in order to obtain
the photon field  (coupling only to the electron),
the  left handed and right handed electron currents
have opposite sign instead of the same signs 
as needed for a pure vector interaction.

(One may try to cure the problem of the signs of terms in the SU(3)   
hypercharge current  by choosing as third member  of the fundamental
triplet instead of the $e_R$ the charge conjugate of the right-handed electron, 
a left handed positron.
This solves the problem of signs.
But causes other  problems difficult to solve.
Thus another path has been chosen to obtain the correct signs for  the terms 
in the hypercharge  current.)

So far, the  singlet combination $\Lambda_0 $ of the nonet, 
see above, has been left aside.
It shall be mixed with the  interaction  $  \Lambda _8 $
in order to obtain the weak  hypercharge interaction,
where  the new field $  X_{\mu}^{8'}$  corresponds to the  $B_{\mu}$- field of the SM
so that the interaction becomes:\\
\\
$  \Lambda _8' =  \tfrac {g}{ \sqrt{3} } (-\overline{\nu_L}  \gamma^{\mu} \nu_L - \overline{e_L} \gamma^{\mu} e_L
                  -2 \overline{e_R} \gamma^{\mu}  e_R ) \cdot X_{\mu}^{8'}      $\\
\\
This implies   rotating the fields and currents   by an angle $\alpha$.
\begin{eqnarray*}\label{alpha}
       \begin{pmatrix}
                X_{\mu}^8   \\
                 X_{\mu}^0
       \end{pmatrix}  =
       \begin{pmatrix}
               cos \alpha &  sin\alpha\\
               -sin\alpha  &  cos \alpha
       \end{pmatrix} \cdot
                 \begin{pmatrix}
                       X_{\mu}^{8'} \\
                       X_{\mu}^{0'}
                  \end{pmatrix}
\end{eqnarray*}
In order to obtain  the wanted interaction term  $\Lambda _8'$   the equation to be solved is\\
\\
$\Lambda _0 +  \Lambda _8 =   $

$ =  \tfrac {\sqrt{2} g}{ \sqrt{3} }  (\overline{\nu_L} \gamma^{\mu} \nu_L + \overline{e_L} \gamma^{\mu} e_L+ \overline{e_R} \gamma^{\mu} e_R )
                 \cdot (cos  \alpha \cdot  X_{\mu}^{0'}  - sin \alpha \cdot X_{\mu}^{8'} $

$ +  \tfrac {g}{ \sqrt{3} }  (\overline{\nu_L} \gamma^{\mu} \nu_L + \overline{e_L} \gamma^{\mu} e_L -2 \overline{e_R} \gamma^{\mu} e_R )
    \cdot  (cos  \alpha \cdot X_{\mu}^{8'}  + sin \alpha \cdot  X_{\mu}^{0'})  \;\;\;\;\;\;\;\;\;\;\;\; \;\;\;\;\;\;\;\; \;\; \;\; (9) $ \\
$= \Lambda _0' +  \Lambda _8'  = $

      $=  \tfrac {\sqrt{2} g}{ \sqrt{3} }  J^{\mu}_{0'} \cdot X_{\mu}^{0'}
       +  \tfrac {g}{ \sqrt{3} }  (-\overline{\nu_L}  \gamma^{\mu} \nu_L - \overline{e_L}\gamma^{\mu} e_L -2 \overline{e_R}\gamma^{\mu} e_R )
                 \cdot  X_{\mu}^{8'} $\\
\\
The solution of equation (9) is given by  
$ cos \alpha = 1/3 $ and $sin\alpha = \sqrt{8/9}$.\\
\\
The new interaction terms  are:\\

$ \Lambda _8'
   =  \tfrac {g}{ \sqrt{3} } (-\overline{\nu_L}  \gamma^{\mu} \nu_L - \overline{e_L} \gamma^{\mu} e_L -2 \overline{e_R} \gamma^{\mu}  e_R )
     \cdot X_{\mu}^{8'}     $

$ \Lambda _0'
  =   \tfrac {\sqrt{2} g}{ \sqrt{3} }  (\overline{\nu_L} \gamma^{\mu}  \nu_L + \overline{e_L} \gamma^{\mu} e_L - \overline{e_R} \gamma^{\mu}  e_R )
     \cdot X_{\mu} ^{'0}
\;\;\;\;\;\;\;\;\;\;\;\;\;\;\;\;\;\;\;\;\;\;\;\;\;\;\;\;\;\;\;\;\;\;\;\; \;\;\;\;\;\; \;\;\;\;\;\;  (10)$\\
\\
Thus, $ \Lambda _8'$ becomes proportional to $ \Lambda _{U(1)}^{SM}$.
The field $ X_{\mu}^{8'} $ can be identified with $B_{\mu}$- field of  $ \Lambda _{U(1)}^{SM}$.

Four of the nine independent boson fields entering 
the sum of interaction terms, equation (8),
enter as projections onto the four-velocity of the leptons 
($u^{\mu} \cdot X_{mu}^i ,\;\;   i= 4-7 $).
These projections behave effectively like scalar fields.
In the following, they shall be treated like scalar fields
with the names $X^4,X^5, X^6, X^7 $, in order to
show their relation to the  scalar fields of the  SM.

Firstly, the four interaction terms $\Lambda_4$ to   $\Lambda_7$
shall be recomposed
as it was done for the terms   $\Lambda _1, \Lambda_2$.
Defining
$\lambda_{4\pm 5}  = \lambda_4 \pm i \lambda_5 $  and
$    X^{4\pm 5} = (X^4 \pm i  X^5)/ \sqrt{2} $
and correspondingly for the indices 6,7 one obtains:\\
\\
$ \Lambda _{4+5}
= (g / \sqrt{2}) \cdot \bar{\phi} \cdot \lambda_{4+5}  \cdot   \phi \cdot  X^{4-5}
= (\sqrt{2}g)    \cdot  \overline{\nu_L}   e_R \cdot  X^{4-5}$\\
$ \Lambda _{4-5}
= (g / \sqrt{2}) \cdot \bar{\phi} \cdot \lambda_{4-5}    \cdot \phi \cdot  X^{4+5}
= (\sqrt{2}g)   \cdot   \overline{e_R}    \nu_L \cdot  X^{4+5}$\\
$ \Lambda _{6+7}
= (g / \sqrt{2}) \cdot \bar{\phi} \cdot \lambda_{6+7}  \cdot   \phi \cdot  X^{6-7}
= (\sqrt{2}g)    \cdot  \overline{e_L}   e_R \cdot  X^{6-7}$\\
$ \Lambda _{6-7}
= (g / \sqrt{2}) \cdot \bar{\phi} \cdot \lambda_{6-7}    \cdot \phi \cdot  X^{6+7}
= (\sqrt{2}g)   \cdot   \overline{e_R}    e_L \cdot  X^{6+7}$\\
\\
The four resulting fields can be identified with the four
independent fields associated  with the complex isospin doublet of the SM: With 
\begin{eqnarray*}\label{Higgsdoublet}
  \Phi =
           \begin{pmatrix}
              \Phi^+   \\
              \Phi^0
            \end{pmatrix}  \;\; \;\; \;\; \;\; \;\; \;\;
      \Phi^{\dag}   =  ( \Phi^{+*}, \Phi^{0*} )
\end{eqnarray*}
the four corresponding interaction terms of the SM are\\
\\
$f_e ( \overline{\chi_L } e_R \Phi +  \overline{e_R }  \chi_L   \Phi^{\dag})  =
f_e ( \overline{\nu_L}   e_R \cdot \Phi^+   + \overline{e_L}  e_R \cdot \Phi^0
     + \overline{e_R}  \nu_L \cdot \Phi^{+*}+ \overline{e_R}  e_L \cdot \Phi^{0*}) $\\
\\
Thus, the identifications
$ X^{4-5}= \Phi^+ $, $ X^{4+5}= \Phi^{+*}  $, $ X^{6-7}= \Phi^{0} $ and $ X^{6+7}= \Phi^{0*}  $
are obvious.
The coupling constant $f_e$  is another independent  constant of the SM.
In the present approach these coupling constants are determined by the
SU(3) symmetry.

\section{Rotation  towards Light}
After having obtained an interaction term corresponding to that of the Standard Model (SM)
for the hypercharge current interaction  with  the $B$ field, the next
rotation  aims to obtain the field $A_{\mu}$ corresponding to the photon
which couples  only to the charged leptons $e_R, e_L$, more precisley to
$J^{\mu} _{e.m.} =  -(\overline{e_L} \gamma^{\mu} e_L + \overline{e_R} \gamma^{\mu} e_R)   $
and the   new Z boson field $Z_{\mu}$,
by mixing $B_{\mu}$ and $ W_{\mu}^0 $.
This  rotation is  described by  the Weinberg angle.

For the considerations in the following section it is  useful to
define  'bare' currents, where no  quantum number factors are included.
In particular, the three possible bare neutral currents are:\\ 

$ \overrightarrow{j^{\mu}} = (j^{\mu} _1,  j^{\mu} _2,  j^{\mu} _3)   =
( \overline{\nu_L} \gamma^{\mu}  \nu_L,\;\;  
  \overline{e_L} \gamma^{\mu}  e_L,  \;\;
  \overline{e_R} \gamma^{\mu}  e_R$ )\\
\\
The neutral currents of the SM   can then be rewritten as  products of a 3-vector
of quantum numbers 
with the three- vector of these bare  currents:\\

$ J^{\mu}_3  =    \sum  I_{3_i} j^{\mu} _i   
 = \overrightarrow{ I_3} \cdot  \overrightarrow{j^{\mu} } $  with  $\overrightarrow{I_3} = (\tfrac{1}{2 } , -\tfrac{1}{2 } , 0) $

$ J^{\mu} _Y  =    \sum  Y_i j^{\mu} _i       
 =  \overrightarrow{ Y} \cdot  \overrightarrow{j^{\mu} } $   with  $\overrightarrow{Y} = (-1, -1, -2)$

$ J^{\mu} _{e.m.}  =    \sum  Q_i j^{\mu} _i   
 =   \overrightarrow{ Q} \cdot  \overrightarrow{j^{\mu} }  $ with  $\overrightarrow{ Q} = (0, -1, -1)$\\
\\
The   vectors $\overrightarrow{ I_3} $ and $  \overrightarrow{ Y} $,  
as they have been derived based on SU(3),
are, of course, identical to the vectors of
diagonal elements of the generators $\lambda _3  $ and  $\lambda' _8  $
of the group SU(3), apart from a normalization factor.
These 3-vectors  
are orthogonal:\\  

$  \overrightarrow{ I_3} \cdot \overrightarrow{ Y} = 0  $\\
\\
They are also orthogonal to the vector of quantum numbers of the third neutral
current corresponding to the singlet interaction  $\Lambda' _0 $.
The currents   $  J^{\mu} _3 $ and $ J^{\mu} _Y $ are orthogonal in the SM, too.
There,  it is due to  the 
choice of the hypercharge currents 
for the U(1)  part of the SU(2)xU(1) symmetry. 

For a more general approach and in order to facilitate the comparison with the SM,
an SU(3) symmetry breaking shall be introduced, or in other words, 
an additional freedom by allowing the 
$X_{\mu}^{8'} $ field to have a different  coupling constant  ($g'$) from the coupling  ($g$) of  $X_{\mu}^{3} $. 
Thus  the two interaction terms are now, 
identifying $X_{\mu}^{8'}$ with $B_{\mu}$ and $X_{\mu}^{3 }$
with $W_{\mu}^0 $ of the SM:\\

$\Lambda _3  
        = g (\overline{\nu_L} \gamma^{\mu} \nu_L - \overline{e_L} \gamma^{\mu} e_L)  W_{\mu} ^0 
        =2 g  \overrightarrow{ I_3} \cdot  \overrightarrow{j^{\mu}}   W_{\mu} ^0 $

$\Lambda' _8  
  =      \tfrac{g'}{\sqrt{3}} (-\overline{\nu_L} \gamma^{\mu} \nu_L 
                          - \overline{e_L} \gamma^{\mu} e_L -2 \overline{e_R} \gamma^{\mu} e_R ) B_{\mu} 
        =\tfrac{g'}{\sqrt{3}}  \overrightarrow{ Y} \cdot  \overrightarrow{j^{\mu}}  B_{\mu} $\\  
\\  
which are  identical to the two  interaction terms ($\Lambda _3,\Lambda' _8$) 
obtained above,  
except of  the different coupling $g'$ 
assumed for the  second term. 
They coincide with  the neutral   interaction terms of the Standard Model, 
apart from   the already mentioned  factor 2 owing  to different definitions of currents or couplings.
The goal of  rotating  the above fields and currents 
is to obtain the correct photon $A$ interaction with the charged lepton, the electron: 
a purely vector interaction. 
The remaining, orthogonal part of the currents 
will couple  to the other field ($Z$) orthogonal to $A$ obtained after rotation:\\

$\Lambda _A : =  -e \cdot ( \overline{e_L} \gamma^{\mu} e_L + \overline{e_R} \gamma^{\mu} e_R )
                    \cdot A_{\mu}  
           =e J_{e.m.} ^{\mu} A_{\mu}  = e \cdot \overrightarrow{ Q} \cdot  \overrightarrow{j^{\mu} }  $  

$\Lambda _Z  =  g_Z  J_{Z} ^{\mu} Z_{\mu}     $\\
\\
where e is the absolute electric charge of the electron, $e = 1.6.... \cdot 10^{-19}C  $,  
and $g_Z  $  and $J_{Z} ^{\mu}  $  will be  determined together with the correct photon interaction. 
The (inverse) rotation is defined by: 
\begin{eqnarray*}%\label{theta}
       \begin{pmatrix}
               B_{\mu}  \\
               W_{\mu} ^0  
       \end{pmatrix}  = 
       \begin{pmatrix}
               cos \theta &  -sin\theta\\
               +sin\theta  &  cos \theta
       \end{pmatrix} \cdot
                 \begin{pmatrix}
                        A_{\mu} \\
                        Z_{\mu}
                  \end{pmatrix}  
\end{eqnarray*}
Thus,  the following  equation for the sum of interaction terms 
before and after the rotation has to be solved:\\
\\
%\begin{eqnarray}\label{AZ}
$\Lambda _3  + \Lambda _8' 
= g (\overline{\nu_L} \gamma^{\mu} \nu_L - \overline{e_L} \gamma^{\mu} e_L) 
                                   ( cos \theta \cdot  Z_{\mu}   +  sin \theta \cdot A_{\mu} )$   

$\;\;\;\;\;\; \;\;\;\;\;\;\;\;\;\;\;\;
 +  \tfrac{g'}{\sqrt{3}}  (-\overline{\nu_L} \gamma^{\mu} \nu_L
                             -\overline{e_L} \gamma^{\mu} e_L -2 \overline{e_R} \gamma^{\mu} e_R )
                                 ( cos \theta \cdot A_{\mu}   - sin \theta \cdot Z_{\mu}) $

$=:  - e  \cdot (\overline{e_L}\gamma^{\mu} e_L +\overline{e_R}\gamma^{\mu} e_R) \cdot A_{\mu} 
     + g_Z \cdot J_{Z} ^{\mu} \cdot Z_{\mu} 
  =:  \Lambda _A  + \Lambda _Z       $\\
\\
The above equation leads to three constraints:\\ 

$ \overline{\nu_L}\gamma^{\mu} \nu_L \cdot A_{\mu} \cdot (- cos \theta  \cdot  \tfrac{g'}{\sqrt{3}}  +sin \theta \cdot g) = 0 
 \;\;\;\ \;\;\;  \;\;\;\ \;\;\; \;\;\;\ \;\;\; 
   \rightarrow  g' =  \sqrt{3} g \cdot tan \theta \;\;\;\;\;\;\;\; (11)$

$ -2\overline{e_R}\gamma^{\mu} e_R \cdot  A_{\mu} \cdot  cos \theta \cdot  \tfrac{g'}{\sqrt{3}}  
                       =  -e \cdot \overline{e_R} \gamma^{\mu} e_R  A_{\mu}
                  \;\;\;\;\;\;\;\;\;\;\;\;\;\;\; \rightarrow  e =  \tfrac{2g' cos \theta }{\sqrt{3}}  = 2 g \cdot sin \theta  $

$ \overline{e_L}\gamma^{\mu} e_L  \cdot  A_{\mu} ( - cos \theta \cdot  \tfrac{g'}{\sqrt{3}}  -sin \theta \cdot g) 
= -e \cdot \overline{e_L}\gamma^{\mu} e_L  A_{\mu}   
\rightarrow e = g \cdot sin \theta+\tfrac{g' \cdot cos \theta }{\sqrt{3}} =  2 g \cdot sin \theta 
\;  $  \\     
\\
Note, the ratios of all three coupling constants are fixed by the same angle. 
The  new interaction terms  for the photon A and the Z boson  are:\\
\\
$ \Lambda _A = e ( - \overline{e_L} \gamma^{\mu} e_L - \overline{e_R} \gamma^{\mu} e_R )  A_{\mu}
    = e    J_{e.m.} ^{\mu}   A_{\mu} \;\;\;\;\;\;\;\;\;\;\;\;\;\;\; $  
\\
$ \Lambda _Z = g (\overline{\nu_L} \gamma^{\mu} \nu_L 
             - \overline{e_L} \gamma^{\mu} e_L )  Z_{\mu}cos \theta$ 

$\;\;\;\;\;\;\ +  \tfrac{g'}{\sqrt{3}}  (-\overline{\nu_L} \gamma^{\mu} \nu_L - \overline{e_L} \gamma^{\mu} e_L -2 \overline{e_R} 
          \gamma^{\mu} e_R )(- sin \theta \cdot Z_{\mu})  $

$ = \tfrac{e}{2 sin \theta cos \theta }[(\overline{\nu_L} \gamma^{\mu} \nu_L  - \overline{e_L} \gamma^{\mu} e_L)]
     + 2 sin^2 \theta( \overline{e_L} \gamma^{\mu} e_L + \overline{e_R} \gamma^{\mu} e_R )] Z_{\mu}$ 
  
$= \tfrac{e}{sin \theta cos \theta }( J^{\mu}_3  - sin^2 \theta J^{\mu}_{e.m.} )Z_{\mu}   $\\
\\
They are identical to the interaction terms of the SM.
If one now requires in addition that 
the currents coupling to the photon $A$ 
and to the $Z$ boson
are orthogonal as were the two currents $J^{\mu}_3$ and $J^{\mu}_Y$ before the rotation 
for both, the SM and the present SU(3) symmetry approach, 
the three-vectors of corresponding quantum numbers have to be orthogonal:\\

$( \overrightarrow{I_3} - sin^2 \theta  \overrightarrow{Q}) \cdot  \overrightarrow{Q} = 0  \;\;\;\;\;\;  \rightarrow  sin^2 \theta  
=  \overrightarrow{I_3} \cdot   \overrightarrow{Q}/   \overrightarrow{Q} ^2 $.\\ 
\\
The immediate consequence is that \\

$sin \theta = \tfrac{1}{2 }    $  and $cos \theta = \sqrt{3/4}     $ \\
\\
and  the above  three coupling constants which were introduced at the beginning
turn out to be equal:\\

      $e =  g = g' $\\
\\  
The Z boson then  couples in a purely axial -vector interaction to the electron:\\ 
\\
$\Lambda _Z =\tfrac{e}{2 sin \theta cos \theta } (\overline{\nu_L} \gamma^{\mu} \nu_L  - \overline{e_L} \gamma^{\mu} e_L
  + \tfrac{1}{2 } \overline{e_R} \gamma^{\mu} e_R  + \tfrac{1}{2 } \overline{e_L} \gamma^{\mu} e_L  ) Z_{\mu}$  

$  =  \tfrac{e}{\sqrt{3}} (2\overline{\nu_L} \gamma^{\mu} \nu_L  - \overline{e_L} \gamma^{\mu} e_L
  + \overline{e_R} \gamma^{\mu} e_R ) Z_{\mu}   $ \\
\\
This consequence applies to the SM and a fortiori to the  SU(3) symmetric interaction.
However, the tacit condition determining the Weinberg angle was 
that the world consists only of the three leptons. 
It is easy to see, how  the more general orthogonality
relation used in grand unified theories to  predict the Weinberg angle, 
emerges from a generalized orthogonality of currents and corresponding vectors of quantum numbers:\\   
\\
$sin^2 \theta = \sum  (Q_i \cdot (I_3)_i ) / \sum  Q_i^2     $\\
\\
where the sum over i includes all  neutral currents. (See Table 1, last column.)

\section{SU(3) Extended to Quarks }
The  Standard Models %\cite{GSW} 
of particle physics 
(see  \cite{Textbooks} %Perkins, Griffith, Okun,Martin and Shaw, Close, 
and \cite{PDG}) assume that quarks have fractional charges.  
In units of e  the charges  are  
+2/3 for  up, charm and top quarks and -1/3  for down, 
strange and bottom quarks independent of their colour, as proposed by
Gell-Mann  \cite{Gellmann1964} and Zweig \cite{Zweig1964} in the framework of flavor-SU(3) 
and later adopted by Fritzsch, Gell-Mann and Leutwyler in the framework of 
color-SU(3) and QuantumChromoDynamics \cite{FritzschGellmann1971}, see Table \ref{tab2}.   

Obviously, the quarks with fractional charges do not fit into an electroweak SU(3) symmetry.
There are 4 particles and the corresponding currents which have to be included in the interaction.
Proceeding as in the previous chapter for quarks alone and requiring, 
at the end,  othogonaliy between various neutral 
quark currents would lead to a different Weinberg angle of $sin^2 \theta  =  9/20$,  
see Table \ref{tab2}.
Assuming that the generations of leptons and quarks which we know are complete, the Weinberg angle 
obtained for 3 triplets of leptons  and 3x3 quadruplets of quarks is 
$sin^2 \theta  =  3/8$.
This is the value obtained in some grand unification models \cite{GeGl}.    

However, following the proposal of Han-Nambu \cite{HanNambu1965},
which assumes that fractional charges with values of
$\pm 1/3$  and $\pm 2/3$   are the result of  
averaging over the three colours of quarks  which have integer electric charges $\pm 1 $  or 0,
the world appears to be simpler.
%(So one reason to follow the Han-Nambu scheme is Ockhams principle.)
Yet, the extensive discussion of this model in the past 
 \cite{Textbooks,PatiSalam1973,ChengWilczek1974,LipkinRubinstein1978,IijimaJaffe1981,Witten1977,Ferreira}
has not %(yet) 
led to its  acceptance by particle physics.
Certainly  oversimplified one may summarize the present situation as follows:
%As Lipkin\cite{LipkinRubinstein1978}  has argued, 
One cannot distinguish between the fractional quark charges of the Standard Model and
the integer charges of the Han-Nambu scheme as long as quarks are confined.
%Witten\cite{Witten1977} has argued that 
This holds if processes are examined where the transition amplitude is proportional
to the quark charge. However, for processes where the squares of charges enter, one may well distinguish. 
This latter argument is pursued by the few advocates of the Han-Nambu scheme, like Ferreira \cite{Ferreira}.
The consequences of the Han-Nambu assumption of integer quark charges for the Weinberg angle have already been examined 
in a previous unpublished letter   \cite{Faessler}.   

Table \ref{tab3} shows  the %more or less obvious 
tentative assignments of  electroweak quantum numbers
to the quarks of  the Han-Nambu scheme.

\begin{table}[htdp]
\centering
\begin{tabular}{   |p{2.0cm}|p{1.0cm}|p{1.3cm}|p{2.5cm}|p{2.0cm} |p{2.5cm}| } \hline
quarks, color b,g,r &$ Q$ &  $I_3$    &$Y=2(Q-I_3)$&$Q \cdot I_3$& $\sum Q_iI^i_3/\sum Q^2_i$\\
\hline
    $ (u_L)_b$  &    1    &    +1/2   &      1      &   +1/2      &  \\
    $ (d_L)_b$  &    0    &  $\;-1/2$ &      1      &  $\;\; 0$   & \\
    $ (u_R)_b$  &    1    &  $\;\; 0$ &      2      &  $\;\; 0$   & $ 1/4$           \\
    $ (d_R)_b$  &    0    &  $\;\; 0$ &      0      &   $\;\; 0$  &                 \\
\hline
    $ (u_L)_g$  &    1    &    +1/2   &      1      &   +1/2         &  \\
    $ (d_L)_g$  &    0    &  $\;-1/2$ &      1      &   $\;\; 0$  &                    \\
    $ (u_R)_g$  &    1    &  $\;\; 0$ &      2      &   $\;\; 0$  & $ 1/4$           \\
    $ (d_R)_g$  &    0    & $\;\; 0$  &      0      &  $\;\; 0$   &                 \\
\hline
    $ (u_L)_r$  &    0    &    +1/2   &      -1      &  $\;\; 0$  &  \\
    $ (d_L)_r$  &    -1   &  $\;-1/2$ &      -1      &  +1/2      &                 \\
    $ (u_R)_r$  &    0    & $\;\; 0$  &   $\;\; 0$   &  $\;\; 0$  & $ 1/4$           \\
    $ (d_R)_r$  &    -1   & $\;\; 0$  &      -2      &  $\;\; 0$  &                 \\
\hline
\end{tabular}
\caption{Electroweak quantum numbers of Han-Nambu quarks with integer electric charges.
For the  colour r (lowest 4 lines)  
three  quarks have the same electric charges and weak hypercharges as
the three leptons and the sterile quark is the $(u_R)_r $. 
For the other two colours, b and g, charges  and weak hypercharges have  opposite signs 
to those of  the leptons and the sterile quarks are the $(d_R)_b$ and $(d_R)_g$.   
}
\label{tab3}
\end{table}
%It appears that 
One of the four quarks for each colour is 'sterile' with respect to 
electroweak interactions, like the  $ \nu_R $ is amongst the leptons.  
It is invisible  at least to the same extent as the right handed
neutrino, since
all the values of its electroweak quantums numbers ($Q,Y,I_3$) coupling to the fields are 0.

Thus, as for leptons, only a triplet of quarks has to be considered  for a given colour 
of the 3 assumed colours, in the framework of electroweak interactions.  

The last triplet,  that  for the colour r, [($u_L)_r,(d_L)_r, d_R)_r ]$ has the same 
electroweak quantum numbers as the lepton triplet, see Table \ref{tab2}.
So, obviously the  equation (11) of the previous section leads to the same relations
between coupling constants and Weinberg angle. 
The other two quark triplets  
have different quantum numbers.
So it has to be shown, that the same Weinberg angle
results in these cases. 
For that purpose, the three constraints listed in equations (11) 
are written in a more general form,
so that it is seen, how  the quantum numbers  
to be multiplied to the three corresponding  bare currents 
$ (j^{\mu} _1,  j^{\mu} _2,  j^{\mu} _3)$  
enter in that equation.\\

Comparing the factors in front of the photon field $A_{\mu}  $,  
as it was done to obtain  equation (11),
one obtains three equations of the same form, 
one for each of the three bare neutral currents  $ j^{\mu} _i$:\\

$ 2g \cdot  sin \theta \cdot I_3^i  +  \tfrac{g'}{\sqrt{3}} cos \theta \cdot Y_i  =  Q_i   $\\
\\
or in terms of $I_3 ^i $ and $ Q_i $ only:\\

$ 2 I_3^i \cdot (g \cdot sin \theta  - \tfrac{g'}{\sqrt{3}} cos \theta )
   = Q_i \cdot (e-\tfrac{2g'}{\sqrt{3}} cos \theta )   $\\
\\ 
Since for all the fermion triplets  there is one fermion with $Q_i = 0$ and another one with 
$I_3^j = 0$, one obtains from the remaining single equation, where 
both   $Q_k $ and $I_3^k $ are non-zero, the two relations already obtained before:\\

$e  = \tfrac{2g'}{\sqrt{3}} cos \theta \;\;\;\;\;\;\; $   and 
$ \;\;\;\;\;\;\;  g \cdot sin \theta  = \tfrac{g'}{\sqrt{3}} cos \theta $\\
\\
The  third condition is then automatically satisfied.

To determine the Weinberg angle, the additional requirement was to assume that 
all the fermions participating in the electroweak interaction follow the same scheme as the 
known first three generations. 
The orthogonality of the currents coupling to the photon and to the Z boson 
then determines the Weinberg angle.
For the case of Han-Nambu quarks with integer charges 
and the above assignments of weak isospin and 
hypercharges the same Weinberg angle, $  sin \theta  = \tfrac{1}{2} $,  
results  as  for the lepton triplet.
And the  three coupling 
constants turn out to be equal, $g' = g = e$,  starting from SU(2)xU(1) and of course a fortiori, 
starting from SU(3).

%\newpage
\section{Field Energy Density Terms}
In the preceding sections,  
the Lagrangian $\Lag$ for the  three basic fermions has  been implemented 
by the interaction terms of the fermions with the 8+1 fields 
required by local gauge invariance. 
The  energy density terms corresponding to the 5 free vector and 4 effective scalar  fields 
have yet to be added  explicitely  to $\Lag$.

It has been a major challenge for the present work to find a  formulation of   
these  terms for  the  case where  both vector and  scalar fields are required 
by the same local gauge symmetry 
and to show  the similarity with the corresponding terms 
introduced in the Standard Model (SM).
 
The term for the  field energy density in a Lagrangian  $\Lag$,  
which is invariant  under local SU(N) gauge transformations,   
for the $N^2-1 $ fields is:\\  
 
$ - \tfrac{1}{4} F^a_{\mu \nu} F_a^{\mu \nu} \;\;\;  $   where  
$\;\;\;   F^a_{\mu \nu} 
               =  \partial_{\mu} X^a_{\nu} - \partial_{\nu} X^a_ {\mu} +2 g f_{abc}X^b_{\mu}X^c_{\nu} \;\;\;$ and $a= 1$ to $N^2-1 \;\;\; (12)$\\  
\\
The three fields of the SU(2) part of the SM are, with   $a = 1$ to  $3 $ 

$\;\;\;\;\;   F^a_{\mu \nu} 
               =  \partial_{\mu} W^a_{\nu} - \partial_{\nu} W^a_ {\mu} +2 g f_{abc}W^b_{\mu}W^c_{\nu}$\\   
with the SU(2) structure constants $f_{abc} = \epsilon_{abc} $. 
The field tensor  for the U(1) associated  $B$ vector potential of the SM  is 
$  F_{\mu \nu}
 =  \partial_{\mu} B_{\nu} - \partial_{\nu} B_ {\mu} $.
The scalar field energy density terms  for the two complex scalar fields 
introduced in the SM for the  purpose of mass generation seems to have 
no resemblance to the SU(N) expression (12). 
It is written as: 
$(D_{\mu} \Phi )^*(D^{\mu} \Phi ) \;   $
with  the covariant derivative including minimal coupling: 
$D_{\mu}   = \partial_{\mu} - i g \tau^a A^a_{\mu} -ig' B_{\mu}  $. 

For the present SU(3) invariant $\Lag $  
the ansatz (12) will be the basis for all fields,
the true vector fields and the effectively scalar fields.
Consider first the vector fields found for $a = 1,2,3,8$, 
which were identified  with the  four vector fields of the SM.
At first sight, the extension to 8 fields appears to have the problem 
that through  the terms    $  2 g f_{abc}X^b_{\mu}X^c_{\nu}$ 
for some combinations  of $abc $  the fields  enter,
which  have been identified as effectively scalar fields.   
To deal with them,  the following  redefinition  of  field 
components has been applied:
Since the  vector components of these  fields 
entered the interaction terms  only in the form of 
the invariant scalar product 
$ \; \; \;\;\;\; X^a = X^a_{\mu}u ^{\mu}$
and are not observable, 
the new, redefined   vector components of these scalar fields  are taken  as  

$ X^a_{\mu}  = X^a  u _{\mu}$\\
Obviously,  this definition leads to the same scalar product 
$ X^a  = (X^a  u _{\mu}) u^{\mu} $  since the four-velocity is a unit vector.
Summing  the non-zero  terms $  2 g f_{abc}X^b_{\mu}X^c_{\nu}$ 
it is  then  easy to show that all 
combinations cancel, where scalar fields possibly enter.
They happen to enter only as pairs. 
All what remains are the terms considered also in the SM SU(2) Lagrangian.
Take for example the vector field  $X^1_{\mu}$.
The terms $  2 g f_{1bc}X^b_{\mu}X^c_{\nu}$ 
are only fed by the combinations with 
$   f_{123}, f_{132}, f_{147}, f_{174}, f_{165}, f_{156}  $.
The first two terms yield the SM terms. 
All the other combinations where scalar field would mix in, 
cancel each other, since
$ u _{\mu}  u _{\nu} =  u _{\nu}  u _{\mu} $.
(The above redefinition of vector components associated 
with the effectively scalar fields 
simply produces  parallel effective vector potentials. 
Hence the  cross product term is zero).    
Similar results are obtained for the fields  $X^2_{\mu}$ and  $X^3_{\mu}$.   
The field  $X^8_{\mu}$ coupling to the weak hypercharge current 
remains celebatarian (as is the U(1) field $B_{\mu}$ of the SM and  the 
field $X^0_{\mu}$ of SU(3) associated with the singlet current)
since only  terms with $f_{845}, f_{854}, f_{867}, f_{876}  $
are non-zero and cancel each other, 
all of them being associated with scalar fields. 

Thus, for  the four fields found in the present SU(3) model 
to be effective   vector fields 
and to couple to the same currents as the corresponding SM fields, 
the  field energy density terms  in  $\Lag$ 
also coincide with the terms of the SM.
The role of the fifth vector field which couples 
to the modified SU(3) singlet current has yet to be clarified. 
This task is beyond the scope of the present paper.

Using the same ansatz (12) for the  field energy densities
for the effectively scalar field potentials $ X^a_{\mu} $ 
with the index $a= 4,5,6,7$ %\\
the similarity with the ansatz  
$ \;\; (D_{\mu} \Phi )^*(D^{\mu} \Phi ) \;\;   $ 
of the SM was not obvious at all, at first sight.
All of the field tensors 
$  F^a_{\mu \nu}
 =  \partial_{\mu} X^a_{\nu} - \partial_{\nu} X^a_ {\mu} +2 g f_{abc}X^b_{\mu}X^c_{\nu}$
contain 8 terms with non-zero 
 $f_{abc} $. 
Take for instance $a=4$. The non-zero terms are 
$f_{471}, f_{417}, f_{462}, f_{426}, f_{453}, f_{435}, f_{458}, f_{485}  $.
All these terms imply mixtures of effectively scalar 
with vector fields  which  do not  cancel. 
Moreover, the connection  between the antisymmetrical  field  tensor 
for the  vector potential
$  F^a_{\mu \nu}  =  \partial_{\mu} X^a_{\nu} - \partial_{\nu} X^a_ {\mu} $  
and the  field vector  $ \partial_{\mu} X^a $ 
for the effectively scalar potential is not obvious.    
However, first consider the Abelian part: 

$  F^a_{\mu \nu} F_a^{\mu \nu} = (\partial_{\mu} X^a_{\nu} - \partial_{\nu} X^a_ {\mu}) 
\cdot  (\partial^{\mu} X_a^{\nu} - \partial^{\nu} X_a^{\mu})  $\\
for the 'effectively scalar' potential  
where again the vector components for the scalar fields 
are redefined as  above:  
$ X^a_{\mu}  = X^a  u _{\mu}$.
Inserting this definition, 
the above field energy density term in $\Lag$ can be shown  to result in:\\

$ - \tfrac{1}{4} F^a_{\mu \nu} F_a^{\mu \nu} 
= -  \tfrac{1}{2}  \partial_{\mu} X^a  \partial^{\mu} X_a u_{\nu} u^{\nu}  +  \tfrac{1}{2}  \partial_{\mu} X^a \partial^{\nu} X_a  u_{\nu} u^{\mu} 
= -  \tfrac{1}{2}  \partial_{\mu} X^a  \partial^{\mu} X_a   +  \tfrac{1}{2}  d_{\tau} X^a d^{\tau} X_a     $\\
\\
The first term  turns out to be  identical to the energy density 
of scalar fields in the SM.  
There is an additional  term on the right hand side of the above equation. 
It appears like the squared mass  given by  the field projected  to the lepton 4-velocity
and may   be related to the mass term of the SM Higgs fields.
Moreover,  the addition  of the non-Abelian  terms in the product
$F^a_{\mu \nu} F_a^{\mu \nu}  $,  with $  F^a_{\mu \nu}
 =  \partial_{\mu} X^a_{\nu} - \partial_{\nu} X^a_ {\mu} +2 g f_{abc}X^b_{\mu}X^c_{\nu}$
leads  to couplings between the   derivatives of the scalar field
and the vector fields which are very similar to those introduced by the SM.
Thus, next to the free field term
$ \sim \partial_{\mu} X^a  \partial^{\mu} X_a  $ there are terms proportional to $g$, like  
$ \sim g \partial_{\mu} X^a  X_b^{\mu} X_c  $  and $ \sim g d_{\tau} X^a  u_{\mu} X_b^{\mu} X_c  $.
The latter term shows that here also  projections  $  u_{\mu} X_b^{\mu} X_c  $  
of the vector fields for $ b=1,2,3,8  $  
enter. 
The terms  $ \sim g^2 $  contain products like   
$ \sim g^2   X_a  X_b X_c^{\mu} X^d_{\mu} $.
A detailed comparison  of all terms with those of the SM   
is beyond the scope of the  present paper.   
However, an important feature of the 'unified' ansatz 
for  scalar and vector field  energy densities  
presented here and  compared with the SM is 
that it leaves room for the  generation of  masses 
by spontaneous symmetry breaking.
In the SM, the scalar fields are introduced 'by hand' for this purpose.
Here, in the present approach, they follow as natural gauge bosons. 

\section{Summary and Conclusion}
The SU(2) symmetry of the Standard Model for electroweak interactions  
has been extended to SU(3).
Only four of the octet of vector gauge  fields couple 
to the standard lepton currents which conserve chirality.
They can be identified with the four vector bosons of the Standard Model:
The three W - bosons of the Standard Model are directly related 
to the bosons associated with the first three 
generators $\lambda_1,\lambda_2, \lambda_3   $  of present  SU(3).
The fourth SU(3) current,  
associated with the  SU(3) hypercharge generator  $\lambda_8$, 
differs from the weak hypercharge current of U(1) of the SM by the sign of one term.
Mixing with the singlet neutral current is needed 
in order to reproduce the correct weak hypercharge current.
The correct signs of weak hypercharges are needed in order  
to be able to subsequently rotate the neutral (hypercharge $Y$ and $I_3 = 0$)  
currents and corresponding vector fields ($B$ and $W^0$)   such, 
that the photon couples to a purely vector current  of charged leptons.
The need of the  singlet vector gauge field is somewhat peculiar. 
It is not  considered in the  Standard Models, 
neither for electroweak nor strong interactions. 
The question, which role the interaction term ($\Lambda_0'$) plays, is open.

In order to introduce gauge fields, 
which are effectively scalar and couple to currents, 
which do not conserve chirality, 
it is proposed to  replace the mass in the mass term of the Dirac equation by  
the derivative  with respect to the eigentime of the lepton. 
Then  four  formerly 'inactive' vector gauge fields 
couple to chirality changing currents in the 
form of effectively scalar projections of the four-potentials 
of fields  to the four-velocity of the lepton.
They can be identified with the four scalar fields 
introduced in the Standard Model by a complex weak isospin doublet 
for the purpose of mass generation by spontaneous symmetry breaking.
In this sense the scalar fields, 
which are introduced 'ad hoc' in the Standard Model, 
are derived from the SU(3) symmetry of the present approach.

The rotation of the neutral vector fields $B$ and $W^0$  
and corresponding neutral currents such, 
that one of the fields  can be identified with the photon ($A$) 
and the other one with $Z$,   is described  by the  Weinberg angle. 
This angle is  determined by sin$^2\theta  = 0.25$ 
in the present SU(3) symmetric model.
Whilst the photon $A$  is required  
to have  a purely vector coupling to the charged lepton current, 
the resulting $Z$ has a purely axial coupling to them, for this angle.    
However, it is shown that the same angle emerges also from the Standard Model, 
assuming different coupling constants for the currents of SU(2) and of U(1), 
if one assumes in addition, 
that the world consists only of the known leptons. 
Even the coupling   constants  for the SU(2) and U(1) interaction emerge  
to be identical under this assumption.
It is mainly this result, 
which suggests  the notion  of an underlying SU(3) of  the electroweak Standard Model,
for the case of leptons.

It is shown that this  SU(3) symmetry can be extended 
to the known quarks, if the Han-Nambu scheme is adopted, 
which assumes integer or zero charges for quarks, depending on their colour.
The observed fractional charges   1/3 or  or 2/3 are naturally explained   
as averages over three colours.
Averaging is  imposed in the   Standard Model for strong interactions, QCD, by 
the axiom  that only colour singlets are observed in nature as long as 
quarks are confined.   
Despite of the existence of Ockhams  principle, 
the Han-Nambu scheme is not the preferred one of present particle physics. 
The present result may add one more stone in favour of integer quark charges  
without affecting QCD, apart from abandoning the idea, 
that SU(3)$_{colour}$ is an exact symmetry. 
     
The Weinberg angle given by  sin$ ^2\theta  = 0.25$  
is many standard deviations away from the presently best values.
These are  obtained   mainly 
from $Z$-pole observables and neutral-current processes,  
for varying renormalization and other prescriptions, 
see the chapter 'Electroweak Model and Constraints of New Physics' in \cite{PDG}. 
The values are in the range    
sin$^2\theta  =  0.22333$ to  sin$^2 \theta  =  0.23864$ at the Z mass.
However, the value  sin$ ^2 \theta  = 0.25$ is  much closer to the PDG values 
than  predictions from  grand unified models.
The present finding suggests  
that the SU(3) symmetry for electroweak interactions 
is valid at energis not too far above  the $Z$ mass.\\

\noindent \textbf{Acknowledgment}\\
\noindent The support of my colleagues
at the Cluster 'Origin and Structure of the Universe', Munich,
and at CERN, Geneva,                                                            
is gratefully acknowledged. 
In particular, I  thank Dorothee Schaile and  Stephan Paul.
And I thank  my  colleagues Otmar Biebel, Gerhard Buchalla, Andrzej Buras
and Harald Fritzsch for friendly and helpful  conversations.

\end{document}